\documentclass[%
 reprint,
superscriptaddress,
nofootinbib,
 amsmath,amssymb,
 aps,
 pra,
]{revtex4-2}

\usepackage{xcolor} 
\usepackage[version=3]{mhchem} 
\usepackage{graphicx}
\usepackage{dcolumn}
\usepackage{bm}
\usepackage{tabularx}
\usepackage{rotating}
\usepackage{makecell}
\usepackage{listings} 
\usepackage{subfigure}
\usepackage{tikz}
\usepackage{hyperref}

\usepackage{comment}

\usepackage{color,soul}

\begin{document}

\title{Discretization anisotropy in micromagnetic simulations}
\author{Samuel~J.~R.~Holt}
\email{samuel.holt@mpsd.mpg.de}
\affiliation{Max Planck Institute for the Structure and Dynamics of Matter, Luruper Chaussee 149, 22761 Hamburg, Germany}
\affiliation{Center for Free-Electron Laser Science, Luruper Chaussee 149, 22761 Hamburg, Germany}
\author{Andrea~Petrocchi}
\affiliation{Max Planck Institute for the Structure and Dynamics of Matter, Luruper Chaussee 149, 22761 Hamburg, Germany}
\affiliation{Center for Free-Electron Laser Science, Luruper Chaussee 149, 22761 Hamburg, Germany}
\author{Martin~Lang}
\affiliation{Max Planck Institute for the Structure and Dynamics of Matter, Luruper Chaussee 149, 22761 Hamburg, Germany}
\affiliation{Center for Free-Electron Laser Science, Luruper Chaussee 149, 22761 Hamburg, Germany}
\author{Swapneel~A.~Pathak}
\affiliation{Max Planck Institute for the Structure and Dynamics of Matter, Luruper Chaussee 149, 22761 Hamburg, Germany}
\affiliation{Center for Free-Electron Laser Science, Luruper Chaussee 149, 22761 Hamburg, Germany}
\author{Hans~Fangohr}
\affiliation{Max Planck Institute for the Structure and Dynamics of Matter, Luruper Chaussee 149, 22761 Hamburg, Germany}
\affiliation{Center for Free-Electron Laser Science, Luruper Chaussee 149, 22761 Hamburg, Germany}
\affiliation{Faculty of Engineering and Physical Sciences, University of Southampton, Southampton SO17 1BJ, United Kingdom}

\date{\today}

\begin{abstract}
Finite difference based micromagnetic simulations are a powerful tool for the computational investigation of magnetic structures.
In this paper, we demonstrate how the discretization of continuous micromagnetic equations introduces a numerical `discretization anisotropy'.
We demonstrate that, in certain scenarios, this anisotropy operates on an energy scale comparable to that of intrinsic physical phenomena.
Furthermore, we illustrate that selecting appropriate finite difference stencils and minimizing the size of the discretization cells are effective strategies to mitigate discretization anisotropy.
\end{abstract}

\maketitle

\section{Introduction}
Solving differential equations is fundamental to physics, describing a wide range of phenomena from heat transfer to magnetism.
Often, analytical solutions to these equations do not exist; instead, we solve them numerically using computational techniques.
One of the most widely used methods is the finite difference technique, which involves discretizing the differential equations onto a regular lattice.
However, such numerical solutions introduce errors due to this discretization.
Anisotropy arising from the discretization of differential equations onto a regular lattice is a well-known phenomenon~\cite{sescu2015numerical, kumar2004isotropic, donahue1997exchange}.
Despite this, the consequences of discretization anisotropy and strategies to mitigate it are rarely discussed in the context of micromagnetics~\cite{donahue2007micromagnetics}.

Micromagnetics models the physics of magnetic systems using a continuum approximation to represent quantities such as the magnetization, energy density, and effective field.
These approximations take the form of differential equations, which must be discretized onto a mesh to obtain numerical solutions.
The aspect ratio, size, and geometry of the discretization cells can significantly impact the results of these simulations.
Simulations may employ coarse discretization to reduce computational effort; however, this can introduce additional errors, leading to preferred directions and the creation of artificial magnetization structures.
In this study, we focus on the errors introduced by using finite difference approximations in micromagnetic simulations.

\section{Finite Difference Approximations}
In micromagnetics, finite difference approximations discretize the magnetization vector field $\mathbf{m}: \mathbb{R}^3 \to \mathbb{R}^3$ onto a regular grid. Typically, a cuboidal discretization is utilized with grid spacings $(h_\mathrm{x}, h_\mathrm{y}, h_\mathrm{z})$ along each spatial direction. These finite difference techniques transform the continuous micromagnetic equations into discrete equations, which can be numerically solved on the grid to approximate spatial derivatives.

Three-point stencils are commonly used to evaluate these derivatives in widely adopted simulation software such as OOMMF~\cite{donahue1999oommf} and Mumax~\cite{vansteenkiste2014design}. These stencils provide efficient and accurate approximations for derivatives of various orders. Specifically, in the $x$-direction (indicated as $\mathbf{e}_\mathrm{x}$), the three-point stencils for the first and second-order derivatives of the magnetization vector field $\mathbf{m}$ can be expressed at any point $\mathbf{r}=(x,y,z)$ as:
\begin{align*}
    \frac{\partial \mathbf{m}(\mathbf{r})}{\partial x} &= \frac{\mathbf{m}(\mathbf{r}+h_\mathrm{x}\hat{\textbf{x}})-\mathbf{m}(\mathbf{r}-h_\mathrm{x}\hat{\textbf{x}})}{2h_\mathrm{x}} + \mathcal{O}\left( h_\mathrm{x}^2 \right), \\
    \frac{\partial^2 \mathbf{m}(\mathbf{r})}{\partial x^2} &= \frac{\mathbf{m}(\mathbf{r}+h_\mathrm{x}\hat{\textbf{x}})+\mathbf{m}(\mathbf{r}-h_\mathrm{x}\hat{\textbf{x}})-2\mathbf{m}(\mathbf{r})}{h_\mathrm{x}^2} + \mathcal{O}\left( h_\mathrm{x}^2 \right).
\end{align*}
These approximations are second-order accurate.
Higher-order stencils can further improve the precision of derivative approximations. For example, a five-point stencil for the first derivative is fourth-order accurate.

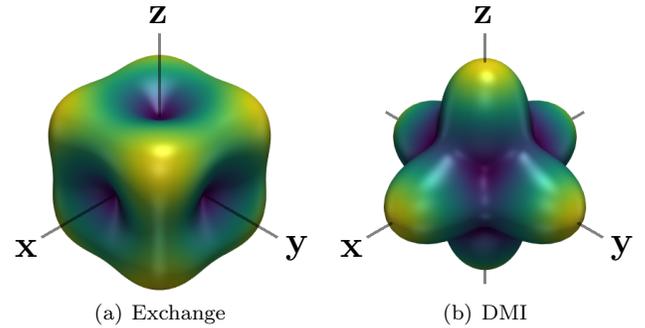
\begin{figure}[t]
\begin{tikzpicture}
        \node[anchor=south west, inner sep=0] (image1) at (0,0) {
            \begin{minipage}{0.5\linewidth}
                \subfigure[Exchange]{ 
                \includegraphics[clip, trim=10cm 7.5cm 10cm 5cm, width=\linewidth]{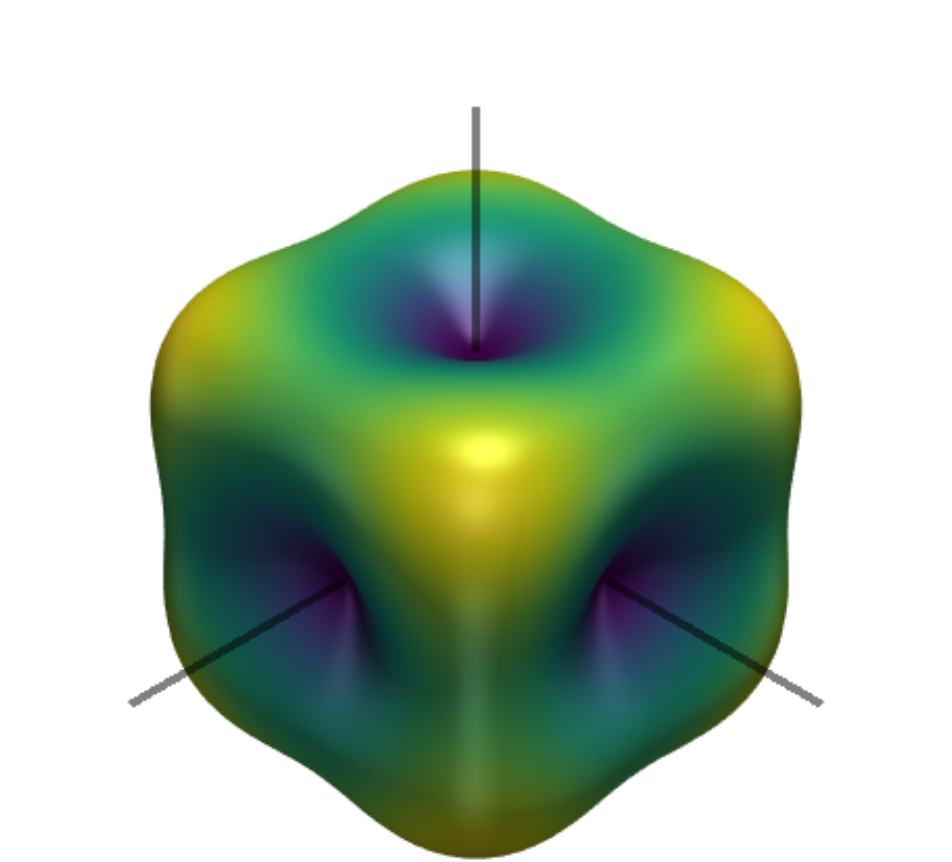}
                }
            \end{minipage}
        };
        \node at (0.5, 1.2) {\Large \textbf{x}}; 
        \node at (4.1, 1.2) {\Large \textbf{y}};
        \node at (2.25, 4.3) {\Large \textbf{z}};

        \begin{scope}[xshift=0.5\linewidth] 
            \node[anchor=south west, inner sep=0] (image2) at (0,0) {
                \begin{minipage}{0.5\linewidth}
                    \subfigure[DMI]{ 
                    \includegraphics[clip, trim=10cm 7.5cm 10cm 5cm, width=\linewidth]{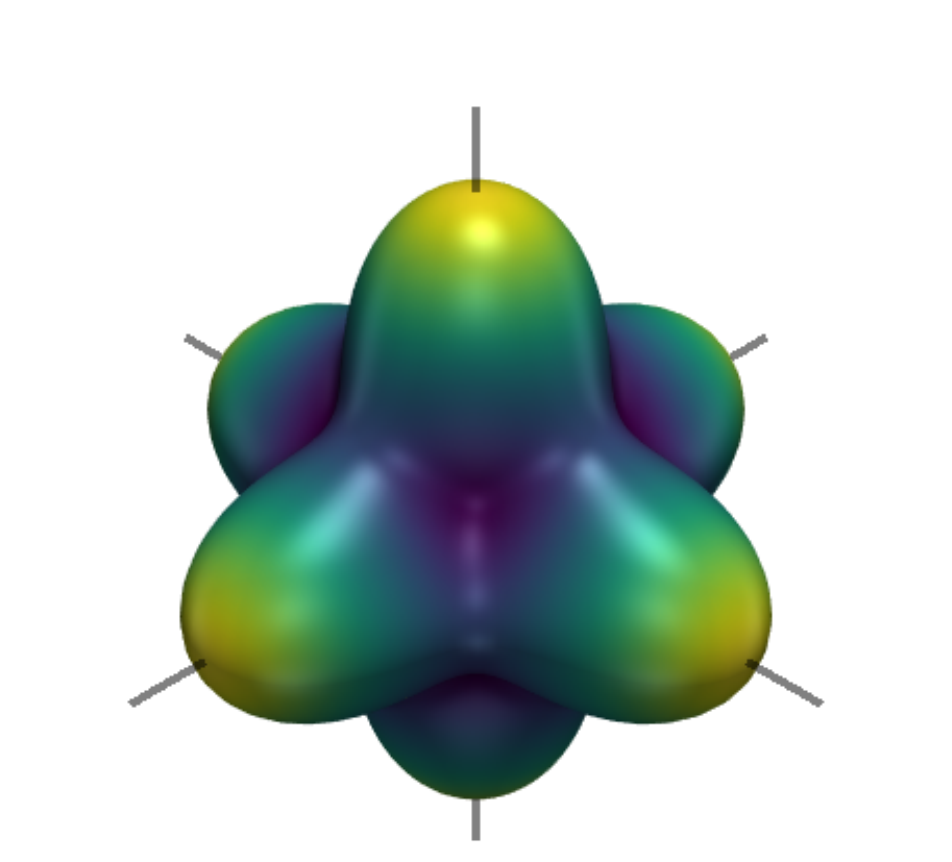}
                    }
                \end{minipage}
            };
        \node at (0.5, 1.2) {\Large \textbf{x}}; 
        \node at (4.1, 1.2) {\Large \textbf{y}};
        \node at (2.25, 4.3) {\Large \textbf{z}};
        \end{scope}
    \end{tikzpicture}
	\caption{Anisotropy of the energy density of (a) exchange, (b) DMI, for a cubic discretization cell with a 3-point finite difference stencil.
  The radial distance from center represents the energy density $w^*$ of a helix propagating in that direction.}
	\label{fig:ani}
\end{figure}

\section{Energy Functionals}
\subsection{Exchange Interaction}
In the continuum approximation, the energy density due to the exchange interaction can be expressed as
\begin{equation}\label{eqn:exchange}
   w_\text{exc} =  - A \mathbf{m}(\mathbf{r}) \cdot \nabla^{2} \mathbf{m}(\mathbf{r}),
\end{equation}
where $A \in \mathbb{R}+$ is the exchange stiffness.

This energy density is isotropic in the continuum model.
To examine this isotropy, we can define a helical ansatz in a helical coordinate system with $\mathbf{\hat{e}_3} \parallel \mathbf{k}$
\begin{equation}\label{eqn:helix}
    \mathbf{m}(\mathbf{r}) = M_{\mathrm{s}}\left(\cos(\mathbf{k}\cdot\mathbf{r})\,\mathbf{\hat{e}_1} + \sin(\mathbf{k}\cdot\mathbf{r})\,\mathbf{\hat{e}_2} + 0\,\mathbf{\hat{e}_3}\right).
\end{equation}
$\mathbf{\hat{e}_1}$, $\mathbf{\hat{e}_2}$, and $\mathbf{\hat{e}_3}$ are orthogonal basis vectors, $\mathbf{k}$ is the wave vector, and $M_{\mathrm{s}}$ is the saturation magnetization.

By substituting \eqref{eqn:helix} into \eqref{eqn:exchange}, the exchange interaction energy density of a magnetic helix in the continuum model reads
\begin{equation}\label{eqn:ex_anal}
    w_\text{exc} = A\mathbf{k}^2.
\end{equation}
This equation is isotropic, and there is no preferential direction for the propagation of the helix.

However, when numerically evaluated using three-point stencils in a finite difference scheme, the energy density takes the form
\begin{equation}\label{eqn:exchange_3point}
\begin{split}
    w_\text{exc}^* =
    & \sum_{\alpha \in \{\mathrm{x}, \mathrm{y}, \mathrm{z}\}} A \frac{2 - 2\cos(k_\alpha h_\alpha)}{h_\alpha^2},
\end{split}
\end{equation}
where $h_\alpha$ is the size of the discretization in the $\alpha$ direction.%
\footnote{If the exchange energy is evaluated using first order derivatives such as 
\begin{equation}
    w_\mathrm{ex} = A \nabla \mathbf{m}(\mathbf{r}) \cdot \nabla \mathbf{m}(\mathbf{r})
\end{equation}
rather than \eqref{eqn:exchange} then the numerical solution is equivalent to \eqref{eqn:exchange_3point} with $h \rightarrow 2h$. 
Even though in the continuum model they are equivalent, the error due to discretisation is different and thus they are only equivalent as  $h \rightarrow 0$.
}

In this discrete form, the energy density becomes anisotropic.

Figure~\ref{fig:ani} (a) depicts the anisotropy for exchange energy density for a cubic discretization cell.
This shows that helices with $\mathbf{k} \parallel \langle 1, 0, 0\rangle$ are energetically favorable, whereas those with $\mathbf{k} \parallel \langle 1, 1, 1\rangle$ are more costly.

\begin{figure}[t]
	\includegraphics[width=\linewidth]{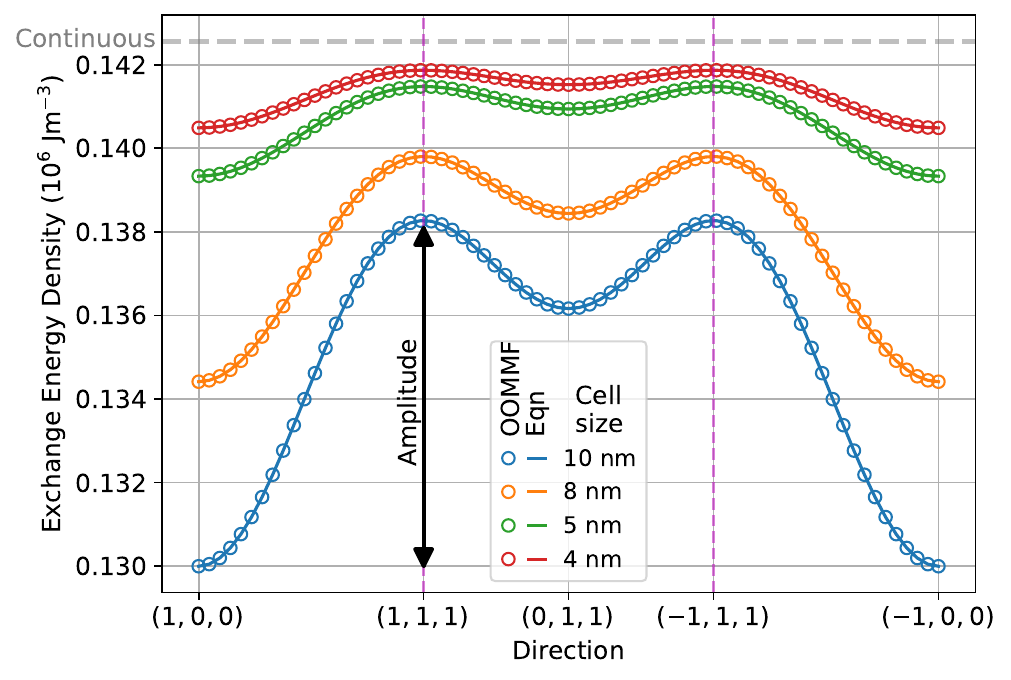}
        \includegraphics[width=\linewidth]{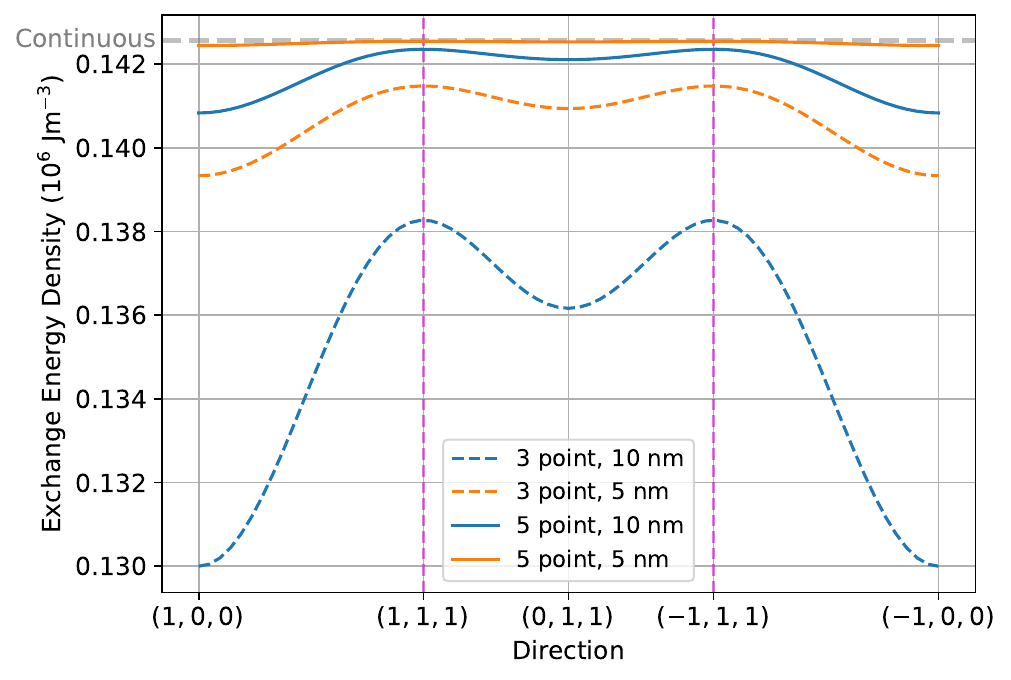}
	\caption{Exchange energy density for Cu$_2$OSeO$_3$ system for a helix of $60\,\mathrm{nm}$ with $\mathbf{k}$ in the $(0,-1,1)$ plane and a cubic discretization cell.
 (Top) The circles show data calculated using OOMMF~\cite{donahue1999oommf, fangohr2024vision} and the solid lines are \eqref{eqn:exchange_3point}.
 The analytical energy as given by \eqref{eqn:ex_anal}.
 (Bottom) Comparison of three point and five-point stencils for the exchange interaction.}
	\label{fig:oommf}
\end{figure}

To quantify the scale of this phenomenon, we introduce a quantity termed the ``amplitude of discretization anisotropy''. 
This represents the difference between the maximum and minimum values of the energy density (due to discretisation). 
The amplitude of discretization anisotropy is labeled in Fig.~\ref{fig:oommf}.
Without discretization, such as for the analytical solution \eqref{eqn:ex_anal}, the discretisation amplitude is zero.

To verify the analytic equation for the three-point-stencil exchange, we compare \eqref{eqn:exchange_3point} with the OOMMF calculator in Ubermag~\cite{donahue1999oommf, beg2021ubermag, fangohr2024vision} in Fig.~\ref{fig:oommf}a using Cu$_2$OSeO$_3$ material parameters and the helical ansatz~\cite{qian2018new}.
OOMMF is a finite difference micromagnetic software which implements three-point-stencils to approximate the exchange interaction.
These equations very accurately describe the numerical implementations of the micromagnetic software.

Using higher-order stencils, such as a five-point stencil, the amplitude of discretization anisotropy decreases while the overall shape remains similar. For instance, with the five-point stencil for the discretized exchange energy density is given by
\begin{equation}
\begin{split}
    w_\text{exc}^* =
    & \sum_{\alpha \in \{\mathrm{x}, \mathrm{y}, \mathrm{z}\}} A \frac{7 - 8\cos(k_\alpha h_\alpha)+\cos^2(k_\alpha h_\alpha)}{3h_\alpha^2}.
\end{split}
\end{equation}
Figure~\ref{fig:oommf}b depicts the improvement of five-point stencils compared to three-point stencils.

\subsection{Dzyaloshinskii–Moriya Interaction}
Similar calculations can also be performed for the Dzyaloshinskii–Moriya interaction (DMI).
In the continuum description, the bulk DMI energy density is given by
\begin{equation}\label{eqn:dmi}
   w_\text{dmi} = D \mathbf{m}(\mathbf{r}) \cdot (\nabla \times \mathbf{m}(\mathbf{r})).
\end{equation}
where $D \in \mathbb{R}$ is the DMI constant.
Evaluating the continuum model with the helical ansatz \eqref{eqn:helix} yields the energy density
\begin{align}
    w_\text{dmi} = -D|\mathbf{k}|.
\end{align}
Like the exchange interaction, this equation is isotropic.

However, when \eqref{eqn:dmi} is discretized and numerically evaluated using three-point stencils, the energy density also becomes anisotropic
\begin{equation}
\begin{split}
    w_\text{dmi}^* =
    & - \sum_{\alpha \in \{\mathrm{x}, \mathrm{y}, \mathrm{z}\}} D \frac{\sin(k_\alpha h_\alpha)}{h_\alpha}\frac{k_\alpha}{|\mathbf{k}|},
\end{split}
\end{equation}

Figure~\ref{fig:ani} (b) depicts the anisotropy of DMI energy density for a cubic discretization cell.
It can be seen that for DMI helices with $\mathbf{k} \parallel \langle 1, 1, 1\rangle$ are energetically favorable. 
Interestingly, this discretization anisotropy is opposite to that observed in the exchange interaction, where helices aligned along the principal axes are more favorable.

Similarly to exchange, using higher-order stencils, reduces the amplitude of the discretization anisotropy.
The equation of the five-point DMI stencil can be produced using the supplementary material.

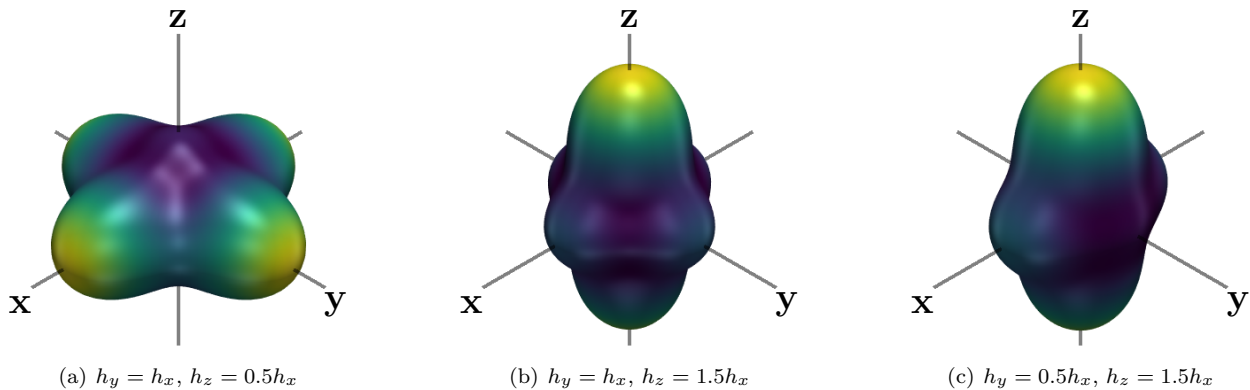
\begin{figure*}[t]
    \centering
    \begin{tikzpicture}
        \node[anchor=south west, inner sep=0] (image1) at (0,0) {
            \begin{minipage}{0.3\textwidth}
                \subfigure[$h_y = h_x$, $h_z = 0.5h_x$]{ 
                \includegraphics[clip, trim=10cm 7.5cm 10cm 5cm, width=\textwidth]{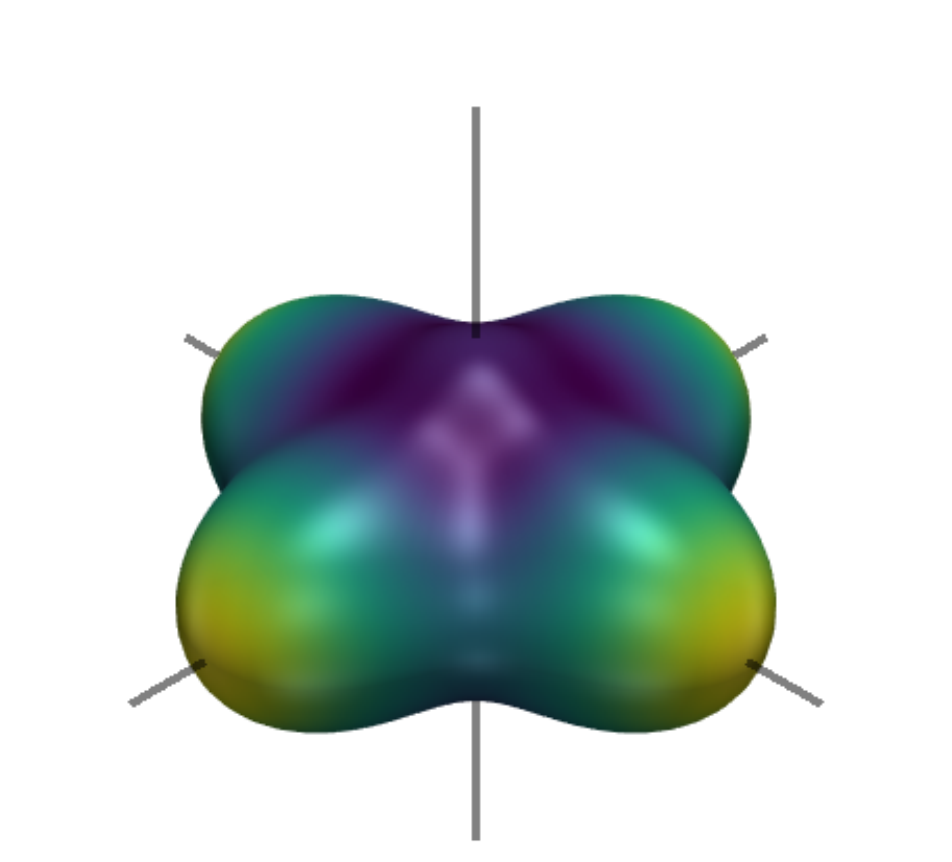}
                \label{subfig:1}
                }
            \end{minipage}
        };
        \node at (0.7, 1.3) {\Large \textbf{x}}; 
        \node at (4.9, 1.3) {\Large \textbf{y}};
        \node at (2.8, 5.1) {\Large \textbf{z}};

        \begin{scope}[xshift=6cm] 
            \node[anchor=south west, inner sep=0] (image2) at (0,0) {
                \begin{minipage}{0.3\textwidth}
                    \subfigure[$h_y = h_x$, $h_z = 1.5h_x$]{ 
                    \includegraphics[clip, trim=10cm 7.5cm 10cm 5cm, width=\textwidth]{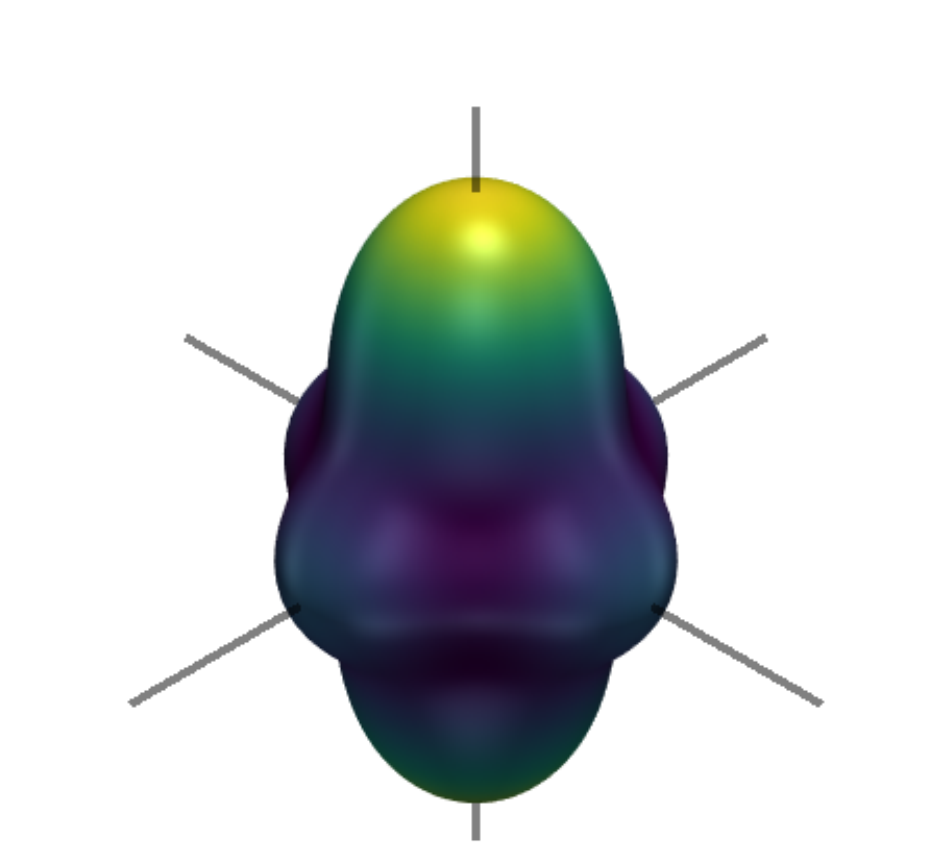}
                    \label{subfig:2}
                    }
                \end{minipage}
            };
            \node at (0.7, 1.3) {\Large \textbf{x}};
            \node at (4.9, 1.3) {\Large \textbf{y}};
            \node at (2.8, 5.1) {\Large \textbf{z}};
        \end{scope}

        \begin{scope}[xshift=12cm] 
            \node[anchor=south west, inner sep=0] (image3) at (0,0) {
                \begin{minipage}{0.3\textwidth}
                    \subfigure[$h_y = 0.5h_x$, $h_z = 1.5h_x$]{ 
                    \includegraphics[clip, trim=10cm 7.5cm 10cm 5cm, width=\textwidth]{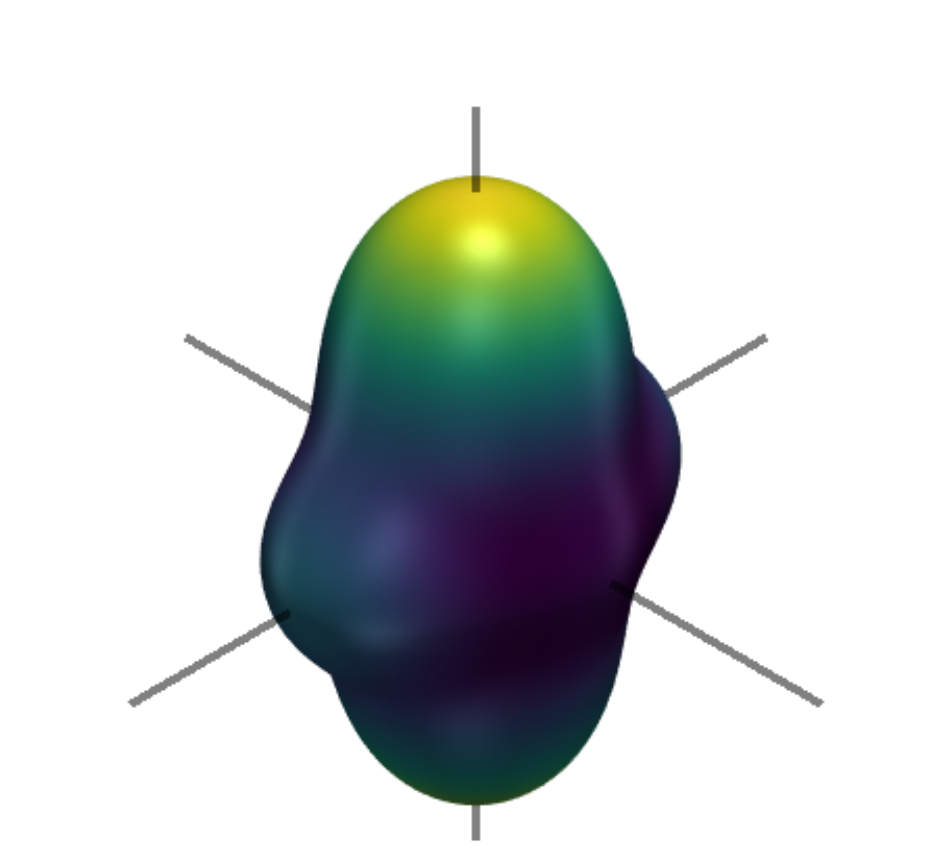}
                    \label{subfig:3}
                    }
                \end{minipage}
            };
            \node at (0.7, 1.3) {\Large \textbf{x}};
            \node at (4.9, 1.3) {\Large \textbf{y}};
            \node at (2.8, 5.1) {\Large \textbf{z}};
        \end{scope}
    \end{tikzpicture}
    \caption{Discretization anisotropy of the DMI energy density, for cuboidal discretization cells with varying aspect ratios for a three-point finite difference stencil.}
    \label{fig:aspect_ratio}
\end{figure*}

\subsection{Other Interactions}
Not all terms commonly present in the micromagnetic Hamiltonian contribute to the discretization anisotropy.
Specifically, it is only introduced by energy terms that involve spatial derivatives of the magnetization vector field $\mathbf{m}$.
For example, terms such as the Zeeman interaction and crystalline anisotropy depend only on the local orientation of the magnetization and therefore do not introduce any discretization anisotropy.
To evaluate the discretization anisotropy arising from energy terms that include spatial derivatives, we first apply finite difference stencils to the Hamiltonian and then substitute a helical ansatz.
We provide Python code in the supplementary information to demonstrate this process, enabling the reader to evaluate and visualize the discretization anisotropy for their own energy terms more conveniently~\cite{DA_SI}.

\subsection{Total Energy}
The total energy density of a system in micromagnetics is the sum of all contributions in the Hamiltonian.
For a system with just exchange and DMI the discretized total energy density reads
\begin{equation}
\begin{split}
    w_\text{tot}^* =
    & \sum_{\alpha \in \{\mathrm{x}, \mathrm{y}, \mathrm{z}\}} A \frac{2 - 2\cos(k_\alpha h_\alpha)}{h_\alpha^2} - D \frac{\sin(k_\alpha h_\alpha)}{h_\alpha}\frac{k_\alpha}{|\mathbf{k}|}.
\end{split}
\end{equation}
Not only is there a competition between the energy terms but there is also a competition between the discretization anisotropies of these terms.

We can calculate the total amplitude of discretization anisotropy for a cubic discretization cell ($h=h_x=h_y=h_z$) 
\begin{equation}
    \frac{h^2|\mathbf{k}|^3}{18} \left( 2D - A|\mathbf{k}|\right) + \mathcal{O}(h^4).
\end{equation}
If this amplitude is positive then the discretization anisotropy due to the DMI dominates and the system favors $\mathbf{k} \parallel \langle 1, 1, 1\rangle$.
Whereas, if this amplitude is negative then the discretization anisotropy due to the exchange dominates and the system favors $\mathbf{k} \parallel \langle 1, 0, 0\rangle$.
The magnitude of the amplitude could be minimized by minimizing the value $2D-A|\mathbf{k}|$, which is however generally not possible as we usually wish to simulate real systems and hence do not have a free choice of any of these variables.
For a system with only exchange and DMI, in equilibrium $|\mathbf{k}| = D/2A$, therefore the discretization anisotropy amplitude will be positive and helices will tend to propagate in a diagonal direction.

In most cases, the best way to minimize the amplitude of discretization anisotropy is minimizing the size of the discretization cell $h$.

Utilizing five-point stencils, for instance, noticeably reduces the anisotropy in the energy density.
The following expression illustrates the reduced amplitude of the anisotropy resulting from the use of a five-point stencil:
\begin{equation}
\begin{split}
    \frac{4h^4|\mathbf{k}|^5}{405} \left( 3D - A|\mathbf{k}|\right) + \mathcal{O}(h^6).
\end{split}
\end{equation}
This expression reveals how the use of more sophisticated stencils can lead to more physically accurate simulations by minimizing the anisotropic effects inherent in simpler discretization approaches.

\begin{figure}[t]
	\includegraphics[width=\linewidth]{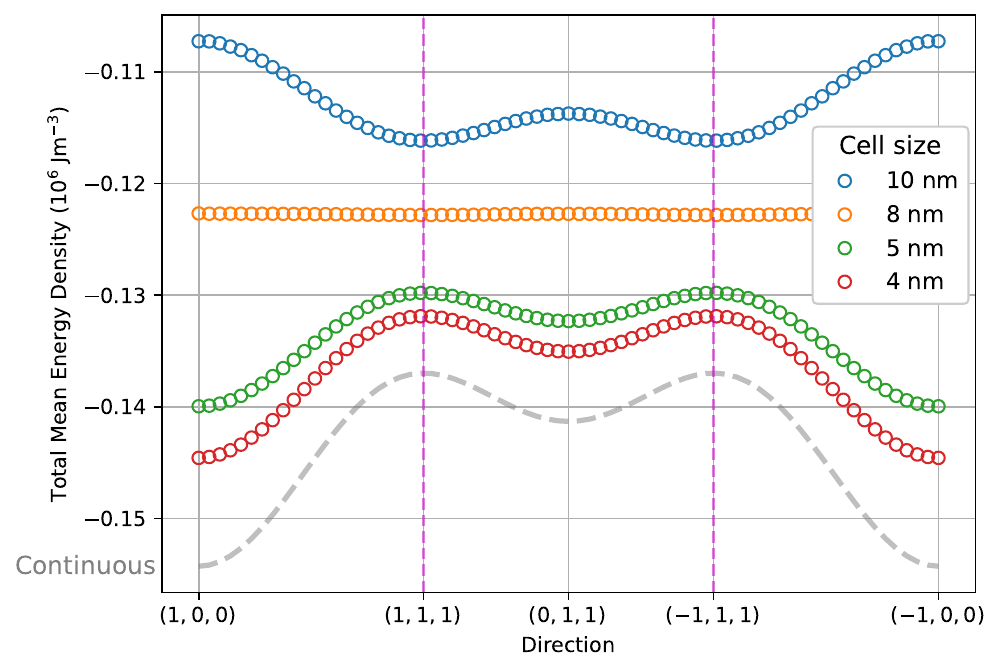}
	\caption{Mean energy density for Cu$_2$OSeO$_3$ system for a helix with $60\,\mathrm{nm}$ with $\mathbf{k}$ in the $(0,-1,1)$ plane and a cubic discretization cell. The anisotropic nature of the cubic anisotropy terms causes the continuous form of the energy density to no longer be a flat line.}
 \label{fig:total}
\end{figure}

\section{Impact}
While many consequences of this numerical phenomenon can be minor, in certain circumstances the discretization anisotropy is of a similar order of magnitude as the physical phenomenon we wish to study. 
For example, Cu$_2$OSeO$_3$ can be modeled with exchange, bulk type DMI, and cubic anisotropy~\cite{qian2018new}.
Whilst exchange and DMI are isotropic, the cubic anisotropy causes the continuous total energy density to become anisotropic with helical state to be favored along the $\langle 1,0,0 \rangle$ directions.
This means rather than a flat line as a function of direction, the continuous solution should be anisotropic.
Figure~\ref{fig:total} shows the average energy density of the system for a helix with $60$\,nm with $\mathbf{k}$ in the $(0,-1,1)$ plane and cubic discretization cells calculated using OOMMF.
For small $h$, the solution approaches the theoretically expected value from the continuous approximation. However, if $h$ is large enough, the characteristics of the system drastically change, leading to a non-physical preferred orientation of the helix along $\langle 1,1,1 \rangle$.

Another critical factor influencing discretization anisotropy is the aspect ratio of the discretization cells.
Figure~\ref{fig:aspect_ratio} shows the discretization anisotropy for the DMI energy using a three-point stencil.
This highlights the preference of the propagation direction of a helix is bias away from gird directions with larger discretization.

\vspace{1em}
\section{Summary}
In summary, the discretization of continuous micromagnetic equations onto a finite difference grid introduces anisotropy.
We prove a helical magnetic state preferentially aligns along certain directions relative to the grid.
However, this methodology can be generalized to other magnetic states. 
These anisotropic effects can be mitigated by selecting appropriate stencils and reducing the grid spacing. 

The impact of discretization anisotropy is wide ranging and includes phenomena such as the energy-minimizing rotation of magnetic structures.
Consequently, in many micromagnetic simulations, magnetic structures can be observed propagating along certain directions even when the systems are isotropic.
More generally, discretization can lead to the formation of nonphysical magnetic structures, often reflecting the underlying symmetries of the discretization.
A thorough understanding of these effects is important for the accurate interpretation of simulation results and for enhancing the overall fidelity of micromagnetic modeling.

\section*{Data availability}
The data and code for this work are available in a public Git repository~\cite{DA_SI}.
We use SymPy~\cite{sympy} for symbolic derivations, with the full setup and analysis scripts detailed in the electronic supplementary materials~\cite{Hunter2007Matplotlib, sullivan2019pyvista, 2020SciPy-NMeth, harris2020array}.

\section*{Acknowledgments}
This project has received funding from the European Union’s Horizon 2020 research and innovation programme under the Marie Skłodowska Curie grant agreement No 101152613 and MaMMoS No 101135546.

%

\end{document}